\documentclass[conference]{IEEEtran}
\IEEEoverridecommandlockouts
\usepackage[numbers]{natbib}
\usepackage{amsmath,amssymb,amsfonts}
\usepackage{algorithm2e}
\usepackage{graphicx}
\usepackage{textcomp}
\usepackage{xcolor}
\usepackage{comment}
\usepackage{booktabs}
\usepackage{subfig}
\usepackage{enumerate}
\usepackage{tikz}
\usepackage{algpseudocode}
\usepackage{listings}
\usepackage{url}
\usepackage{bm}
\usepackage[english]{babel}

\usepackage{scalerel,stackengine}
\stackMath
\newcommand\reallywidehat[1]{%
\savestack{\tmpbox}{\stretchto{%
  \scaleto{%
    \scalerel*[\widthof{\ensuremath{#1}}]{\kern-.6pt\bigwedge\kern-.6pt}%
    {\rule[-\textheight/2]{1ex}{\textheight}}
  }{\textheight}%
}{0.5ex}}%
\stackon[1pt]{#1}{\tmpbox}%
}

\DeclareMathOperator*{\argmax}{arg\,max}

\definecolor{dkgreen}{rgb}{0,0.6,0}
\definecolor{gray}{rgb}{0.5,0.5,0.5}
\definecolor{mauve}{rgb}{0.58,0,0.82}
\newcommand{\bc}{\begin{center}}
\newcommand{\ec}{\end{center}}
\newcommand{\be}{\begin{equation}}
\newcommand{\ee}{\end{equation}}
\newcommand{\bea}{\begin{eqnarray}}
\newcommand{\eea}{\end{eqnarray}}
\newcommand{\beq}{\begin{eqnarray*}}
\newcommand{\eeq}{\end{eqnarray*}}
\newcommand{\bv}{\left( \begin{array}{c} }
\newcommand{\ev}{\end{array} \right) }

\usetikzlibrary{arrows,chains,matrix,positioning,scopes}
\makeatletter
\tikzset{join/.code=\tikzset{after node path={\ifx\tikzchainprevious\pgfutil@empty\else(\tikzchainprevious) edge[every join]#1(\tikzchaincurrent)\fi}}}
\makeatother
\tikzset{>=stealth',every on chain/.append style={join},every join/.style={->}}
\tikzstyle{labeled}=[execute at begin node=$\scriptstyle,execute at end node=$]
\usepackage{enumerate}

\begin{document}

\title{A framework for online investment algorithms}


\author{
\IEEEauthorblockN{A.B. Paskaramoorthy}
\IEEEauthorblockA{ %
\textit{Computer Science and Applied Maths}\\
\textit{University of Witwatersrand} \\
Johannesburg, South Africa\\
andrew.paskaramoorthy@wits.ac.za}
\and
\IEEEauthorblockN{T.L. van Zyl}
\IEEEauthorblockA{ %
\textit{Wits Institute of Data Science} \\
\textit{University of Witwatersrand} \\
Johannesburg, South Africa\\
terence.vanzyl@wits.ac.za}
\and
\IEEEauthorblockN{T.J. Gebbie}
\IEEEauthorblockA{
\textit{Department of Statistical Science} \\
\textit{University of Cape Town}\\
Cape Town, South Africa \\
tim.gebbie@uct.ac.za}
}

\maketitle

\begin{abstract}
	The artificial segmentation of an investment management process into a workflow with silos of offline human operators can restrict silos from collectively and adaptively pursuing a unified optimal investment goal. 
	To meet the investor objectives, an online algorithm can provide an explicit incremental approach that makes sequential updates as data arrives at the process level. 
	This is in stark contrast to offline (or batch) processes that are focused on making component level decisions prior to process level integration. 
	Here we present and report results for an integrated, and online framework for algorithmic portfolio management. This article provides a workflow that can in-turn be embedded into a process-level learning framework.  
	The workflow can be enhanced to refine signal generation and asset class evolution and definitions. Our results confirm that we can use our framework in conjunction with resampling methods to outperform naive market capitalisation benchmarks while making clear the extent of back-test over-fitting. 
	We consider such an online update framework to be a crucial step towards developing intelligent portfolio selection algorithms that integrate financial theory, investor views, and data analysis with process level learning. 
\end{abstract}

\begin{IEEEkeywords}
online adaptive learning, portfolio management, machine learning
\end{IEEEkeywords}

\section{Introduction} \label{sec:intro}

\subsection{Motivation}

The typical quantitative asset management process consists of the preparation of risk and returns predictions, then portfolio optimisation, and lastly performance evaluation. Risk and returns predictions, various constraints, and other decision-making signals are prepared individually and then integrated into the required work-flow\footnote{A particular workflow is a realisation of the required general process that encapsulates a sequence of actions required to achieve an outcome.} to make an investment decision. Investment decisions are then often moderated through the qualitative input of human operators: the fund manager. The decision-making process is inherently serial yet component tasks are executed without full recognition of their dependencies on the modelling decisions of the other parts. 

Most asset management workflows typically pass a spread-sheet (or some equivalent) down a production line. Consequently, the process becomes compartmentalised by virtue of the intermediation of human operators. As a result, workflows, at least in finance, lack sufficient adaption since the update rates are usually orders of magnitude slower that the information flow in actual markets.

Additionally, the compartmentalisation of the components making up the asset management workflow poses a variety of behavioural challenges for system validation and risk management. Human operators can be, both unintentionally and intentionally, incentivised to pursue outcomes that are adverse to the overall achievement of investment goals. For example, this can include engaging in back-test over-fitting of forecast models to improve product marketability, selecting for outcomes that have product supportive narratives, or choosing benchmarks or portfolio performance measures that are easier to out-perform, all often without provable advantage to the investor. The list is seemingly endless and is only limited by the creativity of the agents in each batched silo. 

Historically, the batched formulation of the components of a modularised asset management workflow was necessary for investment decision making in the face of uncertainty.  Segmenting the workflow into compartments had the added benefit of producing specialists for subsets of the workflow; both to mitigate risk, but to also seek an investment advantage. Examples of these legacy perspectives include the segmentation of decisions into active and passive, core and satellite, the use of multi-managers, structured products that exploit the costing of risk transfer, various segmented styles, and the associated financial engineering. These developments have in turn allowed managers to justify fee-layering and the inclusion of various consulting and advisory services that have built on the historically necessary, but perhaps increasingly redundant, batched and compartmentalised approach to asset management. This approach to asset management may no-longer remain best-practice in terms of long-term wealth preservation and growth.

In the era of cost-effective high-performance computation and data-availability, retaining inherently offline \footnote{Here we use batch and offline interchangeably as both use collections, or all of the data; online implementations use a single realisation of features at each algorithm step.} modularisation of the investment workflow can be seen as serving the incentive-structures of the product providers and regulators potentially at the cost of the primary product end-users: pension and mutual fund investors. 

Is there an alternative? Can the entire investment process be transformed to facilitate a machine learning approach to optimal portfolio choice for the investment outcomes in a data starved and wildly random environment? This is an environment where large quantities of new cross-sectional information are always available, but data very quickly becomes redundant as market participants adapt to it. There are precedents in the online trading algorithm literature inspired by the pioneering work of Cover \cite{Cover1991,CO1996}. He introduced a “follow-the-winner” online investment algorithm called the Universal Portfolio (UP) algorithm. Here, the idea was to allocate capital to a set of experts characterised by different portfolios, or trading strategies, and to then let them compete. At each online algorithm update, capital is shifted from losers to winners in proportion to the incremental strategy returns, to find a final aggregate wealth from the consolidated strategy. Subsequent to this a variety of online learning algorithms for trading emerged, of particular importance was the seminal ``Can we beat the best stock?" work by Borodin, El-Yaniv and Gogan \cite{BEG2004}. This cemented the idea that online learning in financial markets was both possible and desirable \cite{li2014online} (See \cite{murphy2019learning} and the references therein for some recent applications for trading on the JSE). 

There are also parallels between these mechanistic online algorithms and linear-quadratic feed-back control and learning frameworks that iterate estimation innovation updates, both from the perspective of the ``Kalman Filter'' (with linear assumptions) - see \cite{Gelb1974} and references therein. As well as more recent advances in reinforcement learning (for weakened and more non-linear assumptions) - see \cite{sutton1998RL} and references therein; for a trading example, see \cite{hendricks2014reinforcement}. Are these viable alternatives? 

The fundamental issue is that of time-scales and information adaption. In financial markets, it may well be that we cannot know the probabilities of the price paths we measure simply because the adaptive processes are changing through time \cite{lo2004adaptive}. The Kalman Filter \cite{Gelb1974}, as an example of the situation where one attempts to retain local linearity to facilitate a simple error feed-back, still requires meaningful knowledge and stability of the underlying model of the processes generating prices. We do not have this. Nor do we currently seem to have enough data to meaningfully machine-learn approximations without generating unacceptable amounts of false positives \cite{R2014}. Current data generation methods including market simulators and resampling methods fail to accurately capture the adaption of price generating processes. Reinforcement learning may prove to be a viable alternative approach if these challenges are overcome~\cite{snow2020machine}. For this reason we argue that we should be guided by the investment process and the data we do have. The problem becomes more a calibration problem aimed at fitting a framework to optimally achieve outcomes, rather than brute-force statistical estimation aimed at statistically selecting models and parameters given the data.

To reiterate: the key issue is the lack of useful data for effective learning and statistical estimation. Historic data quickly becomes redundant in financial markets because of strategic decision making by market participants~\cite{snow2020machine}. For this reason, we take the middle ground and retain the overall portfolio construction process and convert it into an online-update algorithm that can be used for learning. We think that such an online framework for learning investments is a viable alternative; at worst, it may prove to be an intermediate risk-management step towards model-free machine learning for investments~\cite{strehl2006pac}. However, we remain sceptical of model-agnostic or mechanistic online frameworks as they do not retain the traditional components of the asset management process that have emerged with the use of risk management concepts aimed at mitigating the adaptive, inter-related, networked, and complex nature of financial markets \cite{glascher2010states}.

However, in the traditional investment management process, many of the legacy functions, modules, and component-level analyses can be super-ceded by analysis at the process level. For example, model selection can be based on profitability (a {\it process-level decision}), rather than by the calculation of arbitrary p-values thresholds (a {\it component-level decision}). This shifts the emphasis from the statistical validation of individual signals and their associated story {\it e.g.} a particular point-forecast or risk-exposure, to the validation of the entire investment process itself. This is why we convert the investment process to an online framework. 

This has been a fundamental conceptual leap forward in many applications of machine learning. The key point:  We cannot meaningfully determine the validity of the models in isolation from the way they are used for decision making. 


Ideally, we would like an investment process that produces decisions that adapt to changes in the market as information arrives. In the control engineering literature, systems that adapt to new data whilst they are in operation without the interference of a human operator are referred to as \textit{online processes}. For the investment process to become an online process, individual components need to be transformed to contain adaptive mechanisms to automatically update as data becomes available. This will allow decision making to shift from an over-reliance on component level decisions measured in isolation on pathological sequences of past historic data, but inter-mediated by human operators, to process level optimal control monitored by human operators in a way that naturally includes causal relationships in a learning framework. 

The single advantage that machines have in this type of control problem is that they can quickly and consistently exploit a very broad set of features through time with a clear accounting for the variety of hypotheses applied to the data in the context of the entire control problem. This can allow the machines to consistently adapt in a repeatable way to new information as it is revealed, but evaluated and adapted to at the process level in an environment where information is quickly forgotten.

\subsection{Our contributions}

We present an framework to construct investment algorithms to facilitate a machine learning approach to investment. Our framework integrates component decisions of a traditional quantitative investment process across time-slices to allow a process-level analysis of the investment process step-wise through time.

Our approach is to retain the ability to balance the long-term strategic risk-premia with sporadic short-term opportunities that can be used to hedge against regime change, while allowing adaptation in the exposures and investment horizons through time. 

The novelty of our approach is the specification of feed-back mechanisms which adaptively estimates the models and updates the size of portfolio bets according to forecast errors. 

Additionally, we demonstrate how an investment algorithm can be calibrated and validated by adopting principles from machine learning. That is, we specify how to calibrate an algorithm to target out-of-sample performance using cross-validation and we demonstrate how to determine performance estimates from validation samples, accounting for the multitude of independent trials. 

What makes our approach potentially viable in tackling the investment problem is that we do not focus on the statistical validity of individual estimated signals prior to integration into the investment decision making process, nor do we artificially segment the investment opportunity set and the associated investment signals. We do not rely on high-complexity batch-based machine learning approaches that open investment decision making to extreme forms of over-fitting past history. Our framework retains the main components of a traditional quantitative investment strategy so that decisions are interpretable and domain-knowledge can be easily integrated.

The rest of this paper is organised as follows. In section \ref{sec:QPM}, we provide an overview of an investment management process which underlies our framework. In section \ref{sec:OAPM} we present an online estimation framework for portfolio management. In section \ref{sec:results} we provide an empirical demonstration of framework by calibrating and testing an investment strategy. In our conclusion, we indicate shortcomings and future directions. 


\section{Quantitative Portfolio Management}\label{sec:QPM}

Industrial quantitative portfolio management processes are based on Mean-Variance Portfolio Theory (MVPT) \cite{markowitz1952portfolio}. A mean-variance optimal portfolio can be constructed by solving the following optimisation problem:
\begin{eqnarray}
&\argmax_{\mathbf{w}} \left\lbrace \mathbf{w}^{\mathsf{T}}\mathbf{\mu} - \frac{\gamma}{2}\mathbf{w}^{\mathsf{T}}\mathbf{\Sigma} \mathbf{w}  \ \vert \ \mathbf{w}^{\mathsf{T}}\mathbf{1} = 1 \right \rbrace.
\end{eqnarray}
This has optimal portfolio controls as a weight vector $\mathbf{w}^*$ found from a sample covariance $\Sigma$ and sample mean excess returns $\mu$:
\begin{eqnarray}
\mathbf{w}^{*} = \frac{\mathbf{\Sigma}^{-1}\mathbf{1}}{\mathbf{1}\mathbf{\Sigma}^{-1}\mathbf{1}} + \frac{1}{\gamma} \mathbf{\Sigma} \left(\mu -   \mathbf{1} \frac{\mathbf{1}^\mathsf{T} \Sigma^{-1} \mu }{\mathbf{1}^\mathsf{T} \Sigma^{-1}\mathbf{1}} \right).
\end{eqnarray}

Under MVPT portfolio construction is a two-step process. In the first step, an investor uses historical data to forecast the means and covariances of asset returns in a market. In the second step, the investor combines the mean and covariance forecasts to construct an optimal portfolio subject to the investor's constraints. This process can be generalised to more complex choices of moments and risk-measures; however, these modifications do not change the underlying two-step approach to single-period portfolio choice (see for examples \cite{lee2000theory} and references therein). As time passes and more data is accumulated, forecasts are revised and portfolios are reconstructed. Revision is necessary since means and covariances of asset returns are time-varying \cite{cochrane2011presidential}. 

In practice, an investor's portfolio is separated into benchmark and tactical components. When these components are managed separately then the tactical component becomes a form of active management. The separation of a portfolio can be justified within the mean-variance framework by decomposing expected returns $\mu$ into benchmark $\pi$ and active return $\alpha$ components as long as the benchmark is on the efficient frontier. Using
$$\bm{\mu} = \bm{\pi} + \bm{\alpha}$$ an optimal portfolio can be decomposed into benchmark and active components \cite{lee2000theory}: 
\begin{eqnarray}\label{eqn:TAA}
\bm{w} &=& \frac{\mathbf{\Sigma}^{-1}\bm{1}}{\bm{1}^{\mathsf{T}}\mathbf{\Sigma}^{-1}\bm{1}}  +  \dfrac{1}{\gamma}\dfrac{\mathbf{\Sigma}^{-1}(\bm{\pi}\bm{1}^{\mathsf{T}} - \bm{1}\bm{\pi}^{\mathsf{T}})\mathbf{\Sigma}^{-1}}{\bm{1}^{\mathsf{T}} \mathbf{\Sigma}^{-1}\bm{1}} \nonumber \\
&& +  \dfrac{1}{\gamma}\dfrac{\mathbf{\Sigma}^{-1}(\bm{\alpha}\bm{1}^{\mathsf{T}} - \bm{1}\bm{\alpha}^{\mathsf{T}})\mathbf{\Sigma}^{-1}}{\bm{1}^{\mathsf{T}} \mathbf{\Sigma}^{-1}\bm{1}}
\end{eqnarray} 
However, this is not necessarily a mean-variance optimal portfolio; this problem is known as the ``Roll critique'' \cite{roll1977critique}. First, that the naive separation of the globally mean-variance optimal portfolio into a benchmark and active portfolio can be sub-optimal if the benchmark is below the efficient frontier of the complete investment universe. Second, that in practice one cannot meaningfully define the ``Market Portfolio'', what one can do is define strategic objectives and express these in a long-horizon equilibrium framework \cite{lee2000theory}.  

For a retail investor, whose objective is to maximise average return, the benchmark portfolio is often some approximation of a strategic portfolio that reflects long-term equilibrium expectations - this is often thought of as some sort of ``market portfolio''.  For an institutional investor, a benchmark portfolio can be used to represent fund liabilities, and be the required hedge against adverse changes in the value of liabilities. The active component of the portfolio is used to then earn average returns in excess of this strategic benchmark portfolio. The returns from the active portfolio are often phrased in terms of exploiting market inefficiencies. However, in practice, what is more important is that the active portfolio represents differences in the active manager's views from the strategic views irrespective of whether these are true arbitrages \cite{lee2000theory}. 

This approach to portfolio management is central to the success of portfolio management being able to mitigate against sporadic and potentially violent regime changes through time. Mean-variance optimisation is used for the construction of the strategic and tactical portfolios not only because of its simplicity, but more importantly because the framework has well-understood deficiencies that can be compensated for by the human operator and other means of financial engineering. 




\subsection{Difficulties in portfolio construction}

The tasks of prediction and portfolio optimisation are complicated by a range of factors. The preparation of predictions is complicated by the lack of data relative to the complexity of the data generating process. Financial time series represent a single path in the form of high-dimensional data-set that is generated by a highly noisy system that adapts over time \cite{bailey2012sharpe}. Even if there was enough data, the process adapt on time-scales that make system identification almost intractable. 

Furthermore, market adaptations can be considered to be \textit{adversarial} since predictive signals decay as they are exploited \cite{mclean2016does}. Consequently, we find that a large number of models are compatible with a given data-set \cite{bailey2012sharpe}, these models are prone to over-fitting \cite{harvey2016and}, and models require frequent updating. These issues are especially relevant to the detection of mispriced assets which, if they exist, are temporary in nature.

The difficulties encountered in trying to use mean-variance optimisation to construct a portfolio are well known \cite{kolm201460}. In particular, mean-variance optimisation is known to produce highly concentrated portfolios which are extremely sensitive to changes in estimates. Another practical problem faced by industrial investors is how to determine the size of the active bets \cite{black1992global}, which should depend on active risk constraints, confidence in the bets, and also depend structurally on the choice of benchmark leading to, often sub-optimal, unintentional bets. These issues can be ameliorated through various regularisation methods, including the use of shrinkage estimators \cite{ledoit2003improved, jorion1986bayes} and portfolio constraints  \cite{jagannathan2003risk}. The key insight in quantitative asset management was to realise that within a mean-variance framework one could combine long-term strategic objectives, with short-term views that are not in equilibrium but built within an equilibrium framework \cite{black1992global}, and that this framework has the additional advantage of ameliorating many of the estimation issues related to inverting co-variances and estimating means. This does not address either: the back-test over-fitting issue, nor the problem of causality (or at least causal awareness at the level of the process outcomes).

A slightly more detailed description of a simplified quantitative investment process could be: 
\begin{enumerate}
	\item {\it Predict time $t+1$ expected returns and their covariances, at time $t$ using a batch of historic data at times $\{ 0,1,\ldots, t \}$}, then
	\item {\it Regularising predictions for time $t+1$}, and
	\item {\it Select an optimal portfolio subject to constraints at time $t$ to be held until time $t+1$.}
	\item {\it Return to step 1.) , and repeat.}
\end{enumerate} 
Modularisation of the workflow arises because of the specialist skills required in each part of the investment process. As a consequence of this separation, the outputs of each component are assessed using component-specific metrics rather than evaluating their impact on meeting investment requirements at the process level. 

For example, shrinkage intensities for mean and covariance prediction are based on optimising predictive performance (rather than investment performance) based on assumptions about the distribution for returns \cite{ledoit2003improved}. Another example is that constraints in the portfolio construction process may prevent the exploitation of active return opportunities \cite{clarke2002portfolio}. 

However, if predictions were prepared with foreknowledge of the portfolio constraints as well as the subsequent tailoring that the predictions will undergo during portfolio construction, more alpha could be transferred to active portfolios \cite{clarke2002portfolio}. The portfolio control problem could then become something that resembles reinforcement learning \cite{charpentier2020reinforcement, sutton1998RL} but without a simple cost function:
\begin{enumerate}
    \item {\it Observe the state of the market at time $t$},
    \item {\it Evaluate time $t$ portfolio decision actions to be taken at time $t$ and held until time $t+1$ given market states},
    \item {\it Take the portfolio decision actions at time $t$},
    \item {\it Measure process level outcomes at time $t+1$; update the hyper-parameters, update state definitions and measurements, and update actions to be taken},
    \item {\it Return to step 1.), and repeat.}
\end{enumerate}
At time $t$ we are not using information from any time prior to time $t$ to be held until time $t+1$ - the decision is online and we evaluate the entire investment process. 

Here we can learn the optimal combination of market state definitions and algorithm hyper-parameter choices to make optimal process level decisions without needing to be concerned with either over-fitting history or misusing component level decision technologies.

\section{Online algorithmic portfolio management}\label{sec:OAPM}

In the previous section, we described how the investment-decision making process is segmented into various components. The segmentation arose from the practical difficulties of managing an institutional portfolio. Although, this segmentation can be beneficial to an investor, the lack of cohesion across component decisions can erode these benefits. 

Decisions along the investment process can be improved by understanding their impact on portfolio performance. We refer to this as a process-level analysis, since only the output of the decision-making process is of concern. A process-level analysis can highlight redundancies and interactions between intermediary decisions, as well as unintended consequences of intermediary incentives. To conduct a process-level analysis, we require a mathematical description of how investment decisions are made, together with an evaluation methodology to assess the future performance of the investment process. 

Mathematical descriptions for components exist within the literature. Our contribution is to bring these components together into a single framework so that their relationships can be analysed. At a process-level, we find that relationships between the data, the components, and the final investment decisions are not analytically tractable. Rather, a computational approach to evaluation is needed.

The machine learning literature contains various tools to analyse decision-making algorithms. These tools are slowly gaining appreciation in the financial community, primarily due to efforts of Bailey and Lopez De Prado (see \cite{RR2014}, \cite{R2014}). In a series of publications, these authors have designed various statistical tools to facilitate the adoption of machine learning in finance problems. Our contribution is to describe how these tools can be used to conduct a process-level analysis for portfolio management problems.

\subsection{A framework for algorithmic portfolio management}

To facilitate a machine learning approach to investment we need a framework that is path-dependent and iteratively allows the update and evaluation of decisions. We retain the main components of a quantitative investment strategy so that decisions are interpretable and domain knowledge can be easily integrated. Accordingly, our framework consists of three components: prediction, regularisation, and portfolio construction to automatically construct benchmark and active portfolios. The novelty of our approach is the online specification of this framework; models are automatically updated as new data becomes available online, and not offline. 

Our proposed framework constructs benchmark and active portfolios by feeding online Bayesian forecasts from an asset pricing model into a mean-variance optimiser. The modelling decisions relating to model selection, determining shrinkage parameters, and setting portfolio constraints form \textit{hyper-parameters} of the algorithm. Hyper-parameters are selected offline during the portfolio back-testing process. Other offline tasks include: data extraction and cleaning, transformation, and selection of the set of signals. These other tasks are not considered further. 

We can represent the decision work-flow in terms of three inter-connected layers of decisions to be made at time $t$ using only information that arrives at time $t$ : I.A. model updates,  II.A. prediction and regularisation updates and III.A. portfolio optimisation updates (See Figure \ref{fig:workflow_words} in Appendix \ref{app:decisionworkflow}). This can in turn be represented in terms the actual update equations themselves: I.B. the system identification update equations, II.B. the prediction update equations, and III.B, the control update equations (See Figure \ref{fig:workflow_symbols} in Appendix \ref{app:theoryworkflow}).

A data-informed approach to portfolio choice can be computationally intensive. Ordinarily, a historical back-test consists of estimating the return of an strategy at each time period in the holdout period. For dynamically constructed portfolios, models must be updated at each period. Making the estimation process online can substantially reduce the computational burden, since only the most recent component output need to be stored in primary memory rather than the entire history. Furthermore, the online estimation avoids various matrix inversions required to estimate regression parameters. 

The asset pricing framework we adopt (See Figure \ref{fig:workflow_symbols} of Appendix \ref{app:theoryworkflow}) is designed around pricing models where returns are generated from a conditional APT factor model that admits temporary mispricings (``alphas'') in a manner that can accommodate {\it ad-hoc} characteristic model departures from some equilibrium \cite{haugen1996commonality, ferson1999conditioning, harvey2000conditional, wilcox2015pricing, wilcox2014hierarchical}. The model is conceptually specified by the unanticipated returns:
\begin{eqnarray} \label{HCrisk}
r_{i,t+1}-\mathbb{E}_t[r_{i,t+1}]=\sum_{p = 1}^{P} \beta_{i,p,t}\left({r_{p,t+1}-\mathbb{E}_t[r_{p,t+1}]}\right)+\epsilon_{i,t+1} 
\end{eqnarray}
where $r_{i,t+1}$ is the return of asset $i$ in excess of the risk-free instrument, $r_{p,t+1}$ is the excess return on the $p$-th risk factor,  $\beta_{i,p,t}$ is the $i$-th asset's exposure to the $p$-th risk factor, and $\epsilon_{i,t+1}$ is a stochastic disturbance term. 
This approach is hierarchical and incorporates both top-down features ({\it e.g.} global externalities) and bottom-up features ({\it e.g.} local and state dependent information) in a frame-work that includes the potential impact of emergence; which can be proxied by statistical clusters, or transient hidden states, whose membership is represented in the stochastic disturbances \cite{wilcox2014hierarchical}. This can admit non-trivial correlations that may be arbitraged out in an appropriate limit. This allows one to combine {\it ad-hoc} and transient information, such as asset specific characteristic based information, with risk-based time series factor based models for asset pricing \cite{ferson1999conditioning,haugen1996commonality,wilcox2015pricing}. 

In the appropriate long-horizon limits:  
\begin{eqnarray}
\mathbb{E}_t[\epsilon_{i,t+1}] = 0, \mbox { and } \nonumber 
\mathbb{E}_t[\epsilon_{i,t+1}r_{p,t+1}] = 0.
\end{eqnarray}
The above equations represent standard moment restrictions for linear regression models to satisfy the necessary long-horizon no-arbitrage assumptions. While our specific implementation utilises Fama and French \cite{fama1993common} risk factors, the framework we develop is consistent with any (rational) factor model. For simplicity we retain the time-dependent or conditional ``beta'' approach but do not implement cluster based corrections that can be implemented with regards to the stochastic noise \cite{wilcox2014hierarchical}. This makes our implementation a conventional model most similar to that of Ferson and Harvey \cite{ferson1999conditioning,wilcox2014hierarchical} but consistent with the approach of Haugen and Baker \cite{haugen1996commonality} within an overall conditional APT framework. Ideally the asset pricing models should be machine learned rather than theoretical specified - we will revisit this issue in latter discussions.

Expected returns are determined as a combination of systematic return premia, consistent with a conditional APT, but also includes a time-varying risk-free excess return, denoted by alpha $\alpha_{i,t}$:
\begin{eqnarray} \label{eq:HCreturn}
\mathbb{E}_t[r_{i,t+1}] &= \alpha_{i,t} + \sum_{p = 1}^{P} \beta_{i,p,t}\mathbb{E}_t[r_{p,t+1}]
\end{eqnarray}
The time-varying conditional factor loadings are given in terms of information variables:
\begin{eqnarray} \label{eq:loadings}
\beta_{i,p,t} &= b^{(0)}_{i,p} + \sum_{m=1}^{M}b^{(1)}_{i,p}\theta_{i,m,t} + \sum_{k=1}^{K}b^{(2)}_{i,p}z_{k,t} \\
\alpha_{i,t} &= a^{(0)}_i + \sum_{m=1}^{M}a^{(1)}_{p}\theta_{i,m,t} + \sum_{k=1}^{K} a^{(2)}_{i,k}z_{k,t}
\end{eqnarray}
where $\theta_{i,m,t}$ represents the $m$-th information variable corresponding to the $i$-th stock ({\it i.e.} bottom-up information), whilst $z_{k,t}$ represents the $k$-th macro-economic variables ({\it i.e } a top-down risk factor). The dependence on the state-variables are determined by fixed parameters $b^{(1)}, b^{(2)}$, whilst a fixed parameter $b^{(0)}$ denotes a time-invariant portion of the risk exposure typical of static linear pricing models; assuming that the state-variables are sufficient. The quantities $a^{(0)}, a^{(1)}, a^{(2)}$ can be similarly defined to model the temporal variation in alpha.

Our framework consists of two adaptive feedback mechanisms to respond to changes in the return generating process. First, an adaptive estimation mechanism is specified through a generic gradient-based optimisation procedure that updates model parameters iteratively using forecast errors (see appendix \ref{app:stateflow}). Second, an adaptive bet sizing mechanism uses forecast errors to make online estimates of model uncertainty, which is used to resize portfolios in the manner designed by Black and Litterman \cite{black1992global, lee2000theory}. We have explicitly adopted a Bayesian view of uncertainty where expected returns have distributions that reflect our current knowledge. While a distributional assumption for the parameters are not explicitly required in Black and Litterman's derivation, their approach is consistent with the assumption that expected returns are normally distributed. Furthermore, we have two models of expected returns, as indicated by the systematic return model and the active return model. As a practical simplification, the covariance matrix of returns is assumed to known - this need not be the case.

Following Black and Litterman's innovative approach to deviations from market equilibrium  \cite{black1992global} we use a mixed estimator \cite{theil1961pure} to blend the expected return models to obtain the composite estimate $\bm{\mu}_{BL}$: 
\begin{equation}
\bm{\mu}_{BL} = \left[\mathbf{\Omega}_{\pi}^{-1} + \mathbf{\Omega}_{\mu}^{-1}\right]^{-1} \left[\mathbf{\Omega}_{\pi}^{-1}\bm{\pi}   + \mathbf{\Omega}_{\mu}^{-1}\bm{\mu}\right]
\end{equation} 
where $\Omega_{\pi}$ and $\Omega_{\mu}$ represent the covariance hyper-parameter of the posterior distributions for the alternative models of the expected returns. The parameters $\bm{\pi}$ and $\bm{\mu}$ are defined as previously.

Since the estimator appears analagous to a convex combination of partial estimates, we can rewrite the mixed estimator as \cite{o2017black}:
\begin{equation}\label{blEret}
\begin{split}
\bm{\mu}_{BL} = \bm{\pi} + \left( \left[ \mathbf{\Omega}_{\pi,t}^{-1} + \mathbf{\Omega}_\mu^{-1}\right] \mathbf{\Omega}_\mu^{-1}\right) \left( \bm{\mu} - \bm{\pi}\right) 
= \bm{\pi} + \Psi \bm{\alpha} 
\end{split}
\end{equation}
where $\Psi = \left( \left[ \mathbf{\Omega}_\pi^{-1} + \mathbf{\Omega}_\mu^{-1}\right] \mathbf{\Omega}_\mu^{-1}\right)$.

The mixed estimate consists of the systematic return estimate and alphas that are scaled depending on the relative uncertainty between the systematic return estimate and the active return estimate. The greater the relative certainty in the systematic return estimate, the more the alpha diminishes. Thus, the mixed estimator has the effect of shrinking the active return estimates to the systematic return estimates.

We substitute these confidence-scaled alphas in place of the original alphas in the active tactical portfolio so that bet-size can be determined according to our confidence in the estimate. In our implementation, the uncertainty matrices corresponding to each return prediction is determined recursively as the exponentially weighted covariance matrix of the realized prediction errors. This means that as signal strength decays, the tactical portfolio bets will become smaller. 
Using the unanticipated return model (\ref{HCrisk}), the conditional\footnote{Note that ``conditional" refers to both conditional on the time t information and assuming that the expected returns are known} covariance matrix has the form:
\begin{equation}
\Sigma_{\vert \pi} = \mathbf{B}^{\mathsf{T}}\Sigma_{f}\mathbf{B} + \Sigma_{\epsilon}
\end{equation}
where $\Sigma$ is the covariance matrix for all assets, $\mathbf{B}$ is the matrix of factor exposures, $\Sigma_{f}$ is the factor covariance matrix , and $\Sigma_{\epsilon}$ is a diagonal matrix of time-varying idiosyncratic variance.

Parameter uncertainty is an alternative source of variation that must be taken into account in the portfolio construction process.
In the case that both the returns and the expectected returns are normally distributed, and the covariance is known, the resulting optimal portfolio rule is the same as the Markowitz rule, except that the covariance matrix is replaced with the marginal covariance matrix. Thus, the covariance matrix must change to $\mathbf{\Sigma} + \mathbf{\Omega}$ to account for uncertainty in expected returns \cite{meucci2010black}. In our implementation, the covariance matrix applied to strategic portfolio would be adjusted to account for strategic return estimate uncertainty, whilst the active tactical portfolio would use a covariance matrix adjusted by the uncertainty of the mixed active return estimate.

\subsection{Calibration and Evaluation}

The process of calibrating an algorithm involves using data to select hyper-parameters. The objective function can be specified by assuming the density function that generates returns \footnote{For instance, Ledoit and Wolf \cite{ledoit2003improved} assume that returns are generated by a multivariate normal distribution, to derive how the shrinkage intensity should be estimated from data} and maximised computationally.

Alternatively, the investor can avoid making assumptions about the data-generating process by directly evaluating a proxy to the out-of-sample performance of the objective function using historical data. Where assumptions seem unavoidable, historical data can be used to try to directly verify their effectiveness for decision making. Cross-validation, a data resampling procedure common in machine learning, can then be used to form more accurate estimates of future performance than the traditional training-holdout splits used in classical statistical modelling procedures \cite{R2016}. Following Ban {\it et al.} \cite{ban2018machine}, we adopt this data-informed approach to evaluating our objective functions for choosing optimal hyper-parameters. 

At a minimum, our framework requires that values for the following hyper-parameters; these must be selected for a given portfolio management algorithm:
\begin{enumerate}
	\item \textbf{Feature Selection}: Here we select factors that may determine strategic returns and information that can influence the active returns\footnote{These can be combination of risk factors $r_{p,t}$, bottom-up information variables, $\theta_{i,m,t}$, and top-down economic or strategic variables $z_{k,t}$ (see equations \ref{eq:HCreturn} and \ref{eq:loadings}). However, more model agnostic approaches to feature selection are equally (if not more) viable.};
	\item \textbf{Investment Universe Selection}: Here we determine which subsets of listed stocks or other assets and instruments can feasibly comprise an investible universe of assets, and asset groups;
	\item \textbf{Estimation and Prediction Filters:} We then choose appropriate online estimation and prediction procedures for the assumed data-generating process using the selected features and investible universe;
	\item \textbf{Prediction and Rebalancing Frequency}: We select how often models and predictions are to be updated, and over what horizons expectations are formed, and decisions are to be made;
	\item \textbf{Portfolio Choice}: Finally, we set the choice of constraints and adjustments for trading costs given an objective function, the features, investible universe, horizons and modelling framework.
\end{enumerate} 
Various choices, such as the choice of estimation methods, may introduce additional hyper-parameters. For example, in our implementation, we use an adaptive filter to estimate models\footnote{This is an online equivalent to the offline, or batch regression that is conventionally used.}, which requires that we specify a memory parameter which controls the adaptation rate.

As we increase the effective number of the hyper-parameter configurations that are tested, it becomes increasingly likely that the maximal performing configuration is a false discovery with an inflated performance estimate. Care must be taken to adjust performance estimates in light of the multiple testing bias arising from evaluating multiple hyper-parameter configurations on the same dataset. From the literature we have selected two adjusted performance measures. First, the Haircut Sharpe Ratio (HSR) \cite{harvey2015backtesting}. Second, the Deflated Sharpe Ratio (DSR) \cite{bailey2014deflated, AA2019}. As mentioned earlier, a key advantage of an algorithmic approach to calibration is that the number of alternative hypotheses is clearly and necessarily represented. 

\section{Strategy results and implementation} \label{sec:results}

The strategy we implement attempts to take advantage of the differences in asset pricing ability between using characteristics models and risk factor models, as originally reported in Daniel and Titman \cite{daniel1997evidence} and replicated in the South African market \cite{van2003explaining, gebbie2007evidence}. Firm characteristics are used to model active returns and are estimated at each time period using cross-sectional regressions. Risk factors are used to model risk-premia and are estimated using time series regressions. 

To make the models online, we adapt standard methods using exponential smoothing. A detailed description of the implementation is included in Appendix \ref{app:stateflow}\footnote{The associated code and data available at 
\url{https://github.com/apaskara/Algo_Invest}.}.

The offline modelling decisions for the online implementation of our indicative strategy are:

\begin{enumerate}
	\item \textbf{Feature Selection}: For indicative purposes we consider two time-series risk factors (as $r_p$'s in Equation \ref{eq:HCreturn}) and four cross-sectional company attributes (as $\theta$'s in Equation \ref{eq:loadings}), these are respectively:
	\begin{enumerate}
	    \item ``Size'' (SMB) and ``Value'' (HML) Fama and French like risk-factors are used for the risk-premia;
	    \item Book-value-to-price (BVTP), market-value (MV), short-term momentum (MOMS), and long-term momentum (MOML), are the company specific attributes used for the characteristic model.
	\end{enumerate}
	\item \textbf{Investment Universe Selection}: The 100 largest JSE listed equity stocks (as defined by their market capitalisation) are selected. However, the investible set at each time period may be less because of missing data;
	\item \textbf{Estimation and Prediction Filters}: Prediction and estimation models are all implemented as online filters: 
	\begin{enumerate}
	    \item Robust Least-M Adaptive (RLMA) filters \cite{RLME} are used to estimate risk-factor loadings;
	    \item Exponentially Weighted Moving Averages (EWMA) and EWMA covariances are used to estimate risk premia, and the systematic risk respectively; and
	    \item EWMA predictors from Robust Least Squares (RLS) cross-sectional regression models are used to model characteristic returns;
	\end{enumerate}
	\item \textbf{Prediction and Rebalancing Frequency}: Here we make weekly model updates and rebalance the portfolio weekly;
	\item \textbf{Portfolio Choice}: Here we select mean-variance (MV) portfolio optimisation with leverage constraints as our portfolio choice framework.
\end{enumerate}

The hyper-parameters in our strategy are given by: Memory factors for long-term ($\lambda_s$) and short-term ($\lambda_a$) return models, shrinkage intensities for the covariance matrices used in benchmark ($\kappa_s$) and active portfolios ($\kappa_a$), benchmark ($\gamma_s$) and active risk tolerances ($\gamma_a$). We evaluate $2700$ hyper-parameter configurations, using four different active return models giving a total of $10800$ parameters.

\subsection{Data description} \label{ssec:data}

The original dataset was provided by iNet-BFA (now iRess) and consists of weekly-sampled data for equities traded on the JSE over the period 14 January 1994 to 21 April 2017. Following the empirical study by Wilcox and Gebbie \cite{wilcox2015pricing}, the following features were selected to model anomalous returns:  book-value-to-price (BVTP), and market cap (MV). Short-term momentum (MOMS) and Long-term momentum (MOML) are engineered features and are calculated as the quarterly return, and the year-on-year return respectively. Size and Value factor mimicking portfolios are constructed to estimate risk-premia. \\

There is a large amount of missing data in the time series of prices as well as the various features. Our approach to missing data is to assume that a stock is not investible for the period that it's price data is missing. This approach allows the investible universe to change at each period and avoids removing stocks without full histories. The absence of data does not prevent the entire cross-section of models from updating.  For each stock, factor model exposures will remain constant over the period where data is missing. Adaptive estimation will continue as soon as data becomes available. Characteristics models are estimated using cross-sectional least squares at each time period. If characteristics data or price data is missing for a stock, it is excluded from the regression. This approach can indeed bias the regression coefficients of characteristics models to reflect stocks with greater liquidity, which would adversely affect the profitability of active positions. 

\subsection{In-Sample and Out-of-Sample Performance}

\begin{figure*}
    \centering
    \subfloat[\textbf{In-Sample Performance}: The left panel shows that the algorithmic strategy outperforms the benchmark strategies on the in-sample data-set. The right panel suggests that the out-performance is driven by the active portfolio. Since the hyper-parameters of the algorithmic strategy were optimised using the in-sample data, the displayed out-performance is expected. The probability of back-test over-fitting is approximately 76\% suggesting that hyper-parameters are over-fit to the in-sample data.
      ]{\resizebox{0.9\textwidth}{!}{\includegraphics{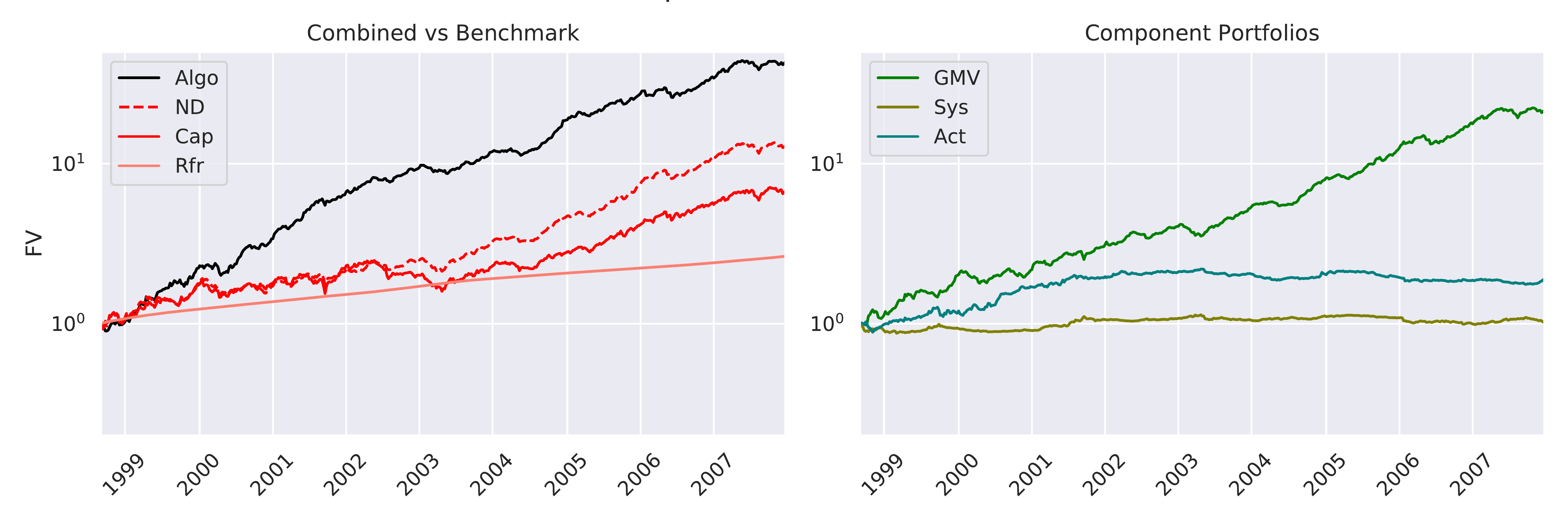}}}
    \newline
    \centering
    \subfloat[\textbf{Out-of-Sample Performance} There is a large deterioration in the out-of-sample performance of all strategies. However, the relative performance degradation of the algorithmic strategy confirms that hyper-parameters were over-fit to the in-sample period. The component performance plots show are indicative of a marked deterioration in active return predictions.  ]{\resizebox{0.9\textwidth}{!}{\includegraphics{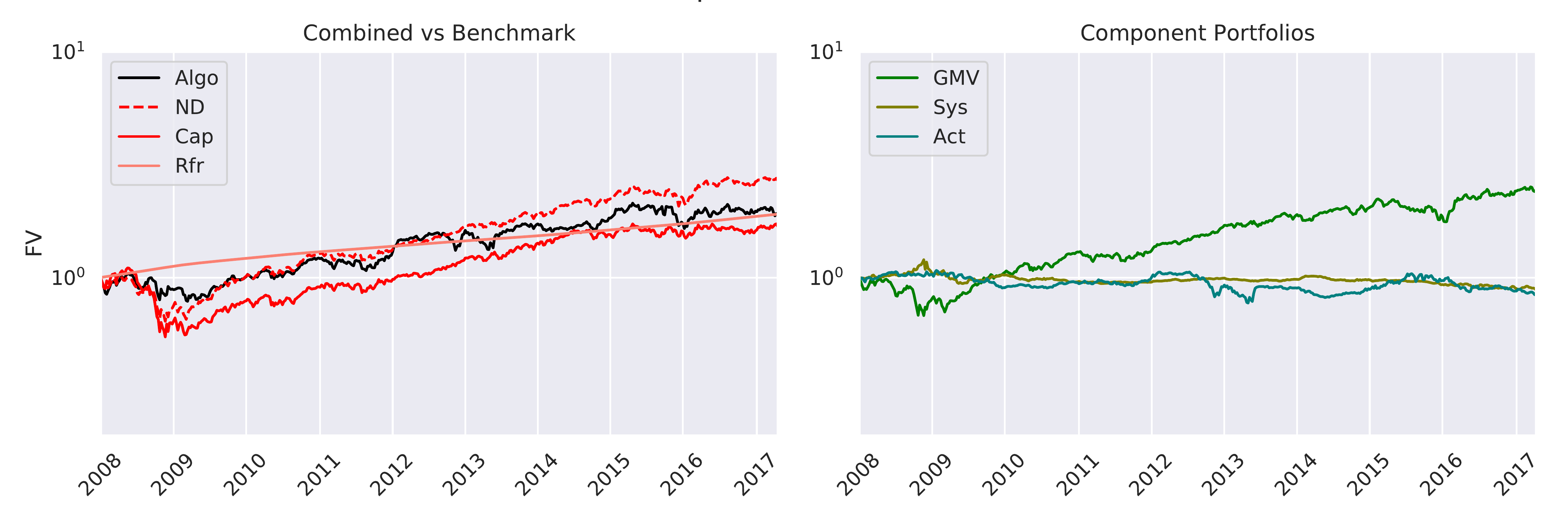}}}
     \caption[Performance]{\textbf{In-Sample vs Out-of-Sample Performance}: The figures in set (a) provide the in-sample (SI) performance, and those in figure set (b), are the out-of-sample (OOS) performance. On the left of each set we compare the cumulative performance of our implemented strategy with the naive-diversification (ND) benchmark, the market-capitalisation weighted (Cap.) benchmark, and the risk-free rate (Rfr). The figures on the right of each set display the performance of the component portfolios: the global minimum variance portfolio (GMV), the systematic bets portfolio (Sys), and the active returns portfolio (Act). Both sets demonstrate the extent to which over-fitting is possible with respect to pre-crises calibration (1994-2007) and post-crises estimation (2008-2017) because of hyper-parameter over-fitting learning is ineffective. Future work is aimed at addressing this key issue.  } \label{fig:perf}
\end{figure*}

In Figure \ref{fig:perf} we provide indicative equity curve simulations and in Table \ref{tab:perf} indicative performance and risk metrics. It is important to realise that although the frame-work is online and adaptive, the hyper-parameters (learning rates,  shrinkage parameters, risk tolerances) still need to be selected prior to using particular operational configuration. In-sample (IS) training is required to select and initialise hyper-parameters. With the selected hyper-parameters the model is run online on the Out-Of-Sample (OOS) test sets.  


\begin{table}[h]
    \centering
    \begin{tabular}{lrrrcrrr}
        \toprule
        \multicolumn{8}{c}{Portfolio Performance}\\
        \toprule
       & \multicolumn{3}{c}{In-Sample} & \phantom{abc} & \multicolumn{3}{c}{Out-of-Sample}  \\
        \cmidrule{2-4} \cmidrule{6-8}
        & SR & DSR & TO && SR & PSR & TO \\
         \midrule
         Algo & 0.3309 & 0.9975 & 0.7165 && 0.0562 & 0.8927 & 0.5680\\
         ND  & 0.2001 & - & 0.1247 && 0.1007 & 0.9831 & 0.0630\\ 
         Cap & 0.1182 & - & 0.0763 &&  0.0480 & 0.8376 & 0.0504\\
         \bottomrule
    \end{tabular} 
    \caption{Model performance over the period of 1994 to 2007 (IS), and 2008 until 2017 (OOS). The Sharpe-Ratio (SR) is weekly and calculated gross of transaction costs. The PSR is calculated against a zero-return benchmark. Turnover (TO) is reported to indicate transaction costs. There is a severe deterioration in out-of-sample performance for the Algorithmic strategy (Algo) indicating that hyper-parameters and offline model choices are overfitting the in-sample period. Interestingly, the in-sample Deflated Sharpe Ratio (DSR) indicates that the trail is a true discovery. The market-cap weighted strategy (Cap) and the naive-diversification strategy (ND) are provided for comparative purposes. }\label{tab:perf}
\end{table}

\begin{figure*}[hbtp!]
\includegraphics[width=0.9\textwidth]{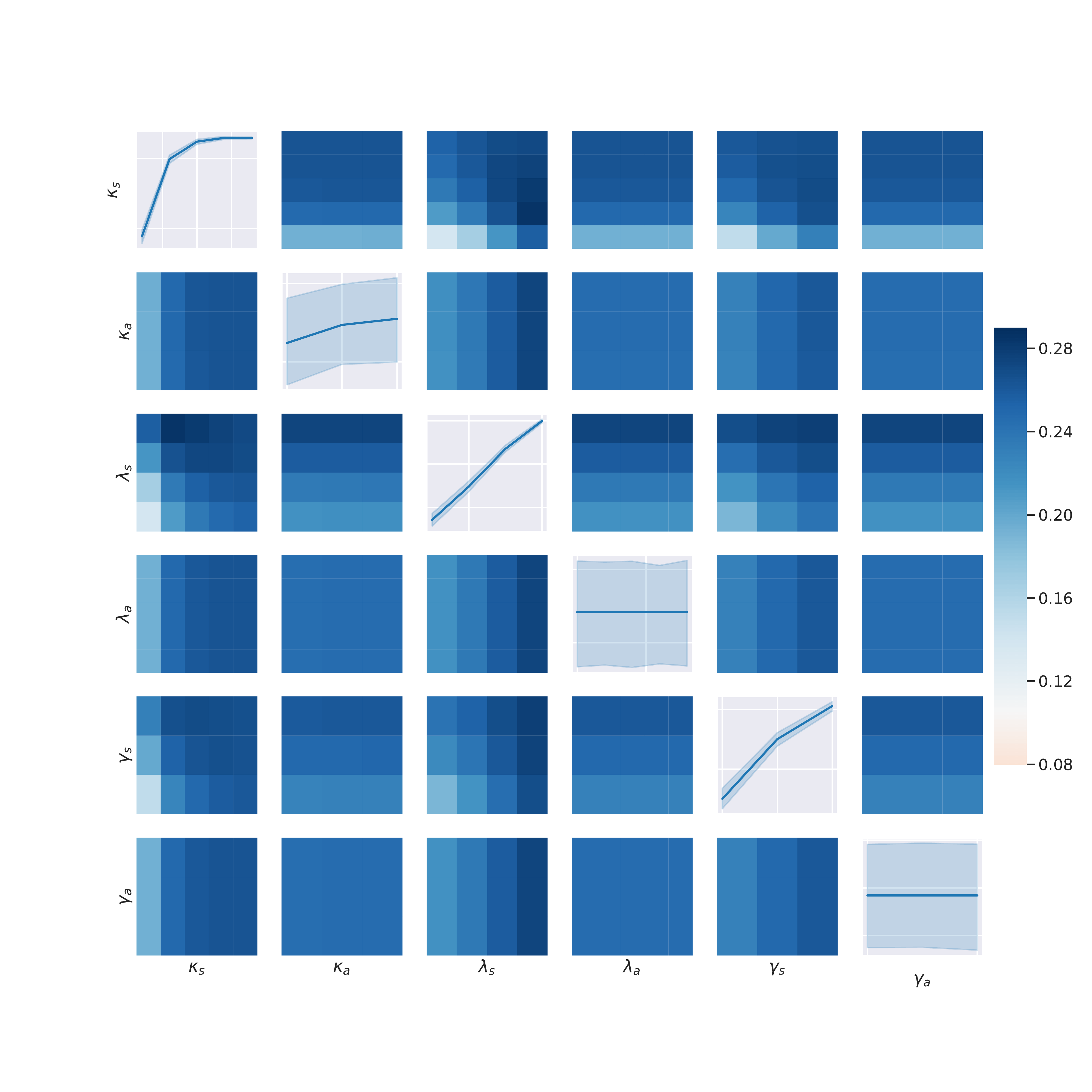}
\caption[Performance]{\textbf{In-Sample Performance for different Hyper-parameter Configurations} The matrix of plots provides a visual depiction of \textit{process-level analysis} by showing how in-sample performance changes with hyper-parameter choice (see Appendix \ref{tbl:Nomenclature} for variable definitions). The main diagonal shows how performance varies for each of the hyper-parameters. The flat slopes in the plots of $\lambda_a$ (the active bet memory factor) and $\gamma_a$ (active bet risk tolerance) indicate that changing these hyper-parameters have minimal effect on in-sample performance. Similarly, for $\kappa_a$, the active bet shrinkage intensity, although there is a slope there is significant dispersion in the response and so the parameter appears degenerate and largely ineffective at the process level. In contrast, the plots for $\kappa_s$, $\lambda_s$, and $\gamma_s$ show that changes in the strategic bet parameters have a greater impact on performance. The off-diagonal plots are heatmaps of in-sample performance for pairs of hyper-parameters and are used to detect interactions. An interactive effect between hyper-parameters may exist if the optimal hyper-parameters in the heatmap are different from the univariate plots.  } \label{fig:hp-perf}
\end{figure*}

We use a traditional 60/40 split to split our data into in-sample and out-of-sample periods. The in-sample period data covers the period from 1994 to 2007 while the out-of-sample period is set to be from 2008 to 2017. Using the in-sample dataset, we use walk-forward cross-validation (also known as time-series cross validation) to select the hyper-parameter configuration producing the largest Sharpe Ratio.  Evaluating the performance of algorithm across the set of potential modelling decisions and hyper-parameter configurations constitutes a form of process-level analysis. In our implementation, the only variable offline decision is the choice of the active return model, which we specify four possibilities. In Figure 
\ref{fig:hp-perf}, we present the estimated performance over the hyper-parameter space for one of the four active return models tested.

Hyper-parameters that maximise the IS Sharpe Ratio are selected for the out-of-sample period. However, the performance estimate given by the maximum IS Sharpe-Ratio is inflated because of the large number of hyper-parameter configurations evaluated. To correct for this, out-of-sample performance is estimated using the Deflated Sharpe Ratio, and is calculated for the Algorithmic Strategy in Table \ref{tab:perf}.


Using the combinatorially symmetric cross-validation procedure \cite{RR2014}, we calculate the probability of backtest overfitting to be $0.7622$. The probability is large, suggesting that it is in fact quite likely that our hyper-parameter selection method detects false strategies. The high PBO is consistent with our finding that out-of-sample performance is substantially worse than in-sample performance as seen in Figure \ref{fig:perf}. 

To further investigate the potential deterioration in performance due to overfitting, we follow the approach of Bailey and Lopez De Prado by regressing the resampled out-of-sample Sharpe-Ratios of the selected strategies onto their corresponding in-sample Sharpe Ratios. The estimated linear equation is given by:
\begin{equation*}
    \mathrm{SR}_{_\mathrm{OOS}} = 0.6340 - 0.9684~ \mathrm{SR}_{_\mathrm{IS}}
\end{equation*}
The negative slope coefficient, together with the $R^2$ of 74\%, is indicative that the active model choice and the selected model are over-fit to the in-sample data set.

\section{Conclusion}

We have presented an indirect adaptive control framework to automate portfolio management that is online. An online process uses data as it arrives and forgets data prior to the current arriving data. An explicitly online representation of the decision process allows one to learn various model and parameter choices to optimally achieve investor outcomes at the process level without the necessary requirement for component level decision variables to aggregate effectively to process level decisions and outcomes. This may allow investment management to migrate from offline and batched human operators integrating decision making without consistent or effect learning to a situation where one can envisage human operators monitoring and risk managing the decision system without unnecessary intervention as the process becomes data-informed and adaptive.  

An investment process typically consists of a long-term benchmark strategy and some sort of mean-reversion strategy that is used to exploit short-term price deviations. Our proposed framework executes both strategies by feeding online Bayesian forecasts from an asset pricing model into a mean-variance optimiser. The novelty of our approach is that the specification is that it is not batch implemented but is explicitly online and updates the entire process incremental as new data-arrives. The feed-back mechanisms which adaptively estimates the models and updates the size of portfolio bets according to forecast errors and realised portfolio performance is made explicit. 

Simulation results confirm that the framework can be used to earn a risk-premia and active-returns whilst adjusting bet size to account for model uncertainty. Empirical tests show that the algorithmic portfolio outperforms the 1/N portfolio and the market portfolio when trading costs are not accounted for. 

We think that our framework is an important  conceptual step towards developing intelligent portfolio selection algorithms that integrate financial theory, investor views, and data analysis in an online workflow rather than staging individual components offline in isolation. The workflow can be enhanced using a more refined approach to signal generation, asset class definitions, and market state extraction. 

The framework we present is an indicative proof-of-principle. The work demonstrates the viability of the workflow with the simplest combination of signals, markets states and asset class definitions that can form a useful investment process. We do not provide a demonstration, or any evidence of a performance advantage. The specific choice of using the in-sample period from 1994 until 2007, {\it i.e} prior to the Global Financial Crises (GFC), and then using the post-crises period as the out-of-sample period, from 2008 until 2017, gives us significant comfort with the negative result arising from our proof-of-principle demonstration. We think it is important to first craft an effective null detection when testing a process level framework. For industrial applications the entire work-flow and systems design would need to be extensively expanded to accommodate mission critical approaches to machine learning, systems management, data-engineering, and effective risk management; while the testing and training periods should be built on more recent data in a manner that leverages algorithm adaption and learning in the quest to avoid valuing adding outcomes where claims of skill cannot be easily differentiated from {\it ex-post} luck.  

\section{Acknowledgement}

The authors would like to thank the attendees at the South African Finance Association (SAFA2020) Conference for their feedback. 

\bibliographystyle{IEEEtran}
\bibliography{IEEEabrv,APTvZTG_SPC_IEEE}

\newpage
\onecolumn
\appendix
\subsection{Model Variables} \label{app:nomenclature}
\begin{table}[h!]
	\centering
	\begin{tabular}{lclc}
		\toprule
		Symbol & \phantom{abc} & Description & \phantom{abc} \\
		\midrule	
		$r_{t}$						&&	Total return 								 \\
		$r^e_t$						&&	Return in excess of the risk-free rate	 \\
		$f_t$						&&	Factor realisations  \\
		$\bm{z}_t$					&&	Information Variables \\
		$\bm{\beta}_t$				&&	Factor Loadings \\
		$\mathbf{B}_t$				&&	Matrix of factor loadings for all stocks  \\
		$\delta_t$					&&	Characteristic payoffs \\
		$\epsilon_t$				&&	Idiosyncratic error  \\
		$\mathbf{\Sigma}_{f,t}$			&&	Covariance matrix of factor realisations \\
		$\mathbf{\Sigma}_{t\vert\pi}$	&&	Conditional covariance matrix  \\
		$\mathbf{\Omega}_{t}$			&&	Expected return uncertainty  \\
		$\mathbf{\Sigma}_{u,t}$			&&	Unconditional covariance matrix  \\
		$\mathbf{\Sigma}_{\epsilon,t}$ 	&&	Idiosyncratic risk matrix \\
		$\mathbf{\Sigma}^{*}_t$ 		&& Shrinkage Target \\
		$\mathbf{\Sigma}_t$ 			&& Risk matrix used in portfolio construction \\	
		$e_t$						&&	Forecast error \\
		$\lambda_{a},\lambda_{s}$					&& Exponential memory factor; (a) active, (s) strategic \\
		$\pi_{t}$					&&	Systematic expected return \\
		$\mu_{t}$					&&	(Characteristic model) expected return \\
		$\alpha_{t+1} $				&& Active expected return  \\
		$\mathbf{\Psi}_t $			&& Active return shrinkage factor \\
		$\alpha_{bl,t}$				&&	Active blended expected return \\
		$\kappa_{a},\kappa_{s}$		&&	Shrinkage Intensity; (a) active, and (s) strategic \\
		$\gamma_{a},\gamma_{s}$		&&	Risk tolerance; (a) active and (s) strategic  \\
		$\bm{w}_t$				&&	Vector of portfolio weights  \\
		$\phi $					&& Predictive function \\
		$C$						&& Portfolio constraint set  \\
		\bottomrule
	\end{tabular}
	\captionsetup{width=0.9\textwidth}
		\caption{Nomenclature for Workflow Charts: The table contains the nomenclature used throughout this article and Figures \ref{fig:workflow_words}  and \ref{fig:workflow_symbols}. The tilde symbol is used for the return and error variables ($\tilde{r_t}$ and $\tilde{e_t}$; see Figures \ref{fig:workflow_words}  and \ref{fig:workflow_symbols}) to denote realisations rather; differentiating from random variables (see Eq. \ref{HCrisk})} \label{tbl:Nomenclature}
\end{table}
\onecolumn
\subsection{Investment Decision Workflow} \label{app:stateflow} \label{app:decisionworkflow}
\begin{figure}[hbtp!]
	\centering
	\includegraphics*[width=0.9\textwidth]{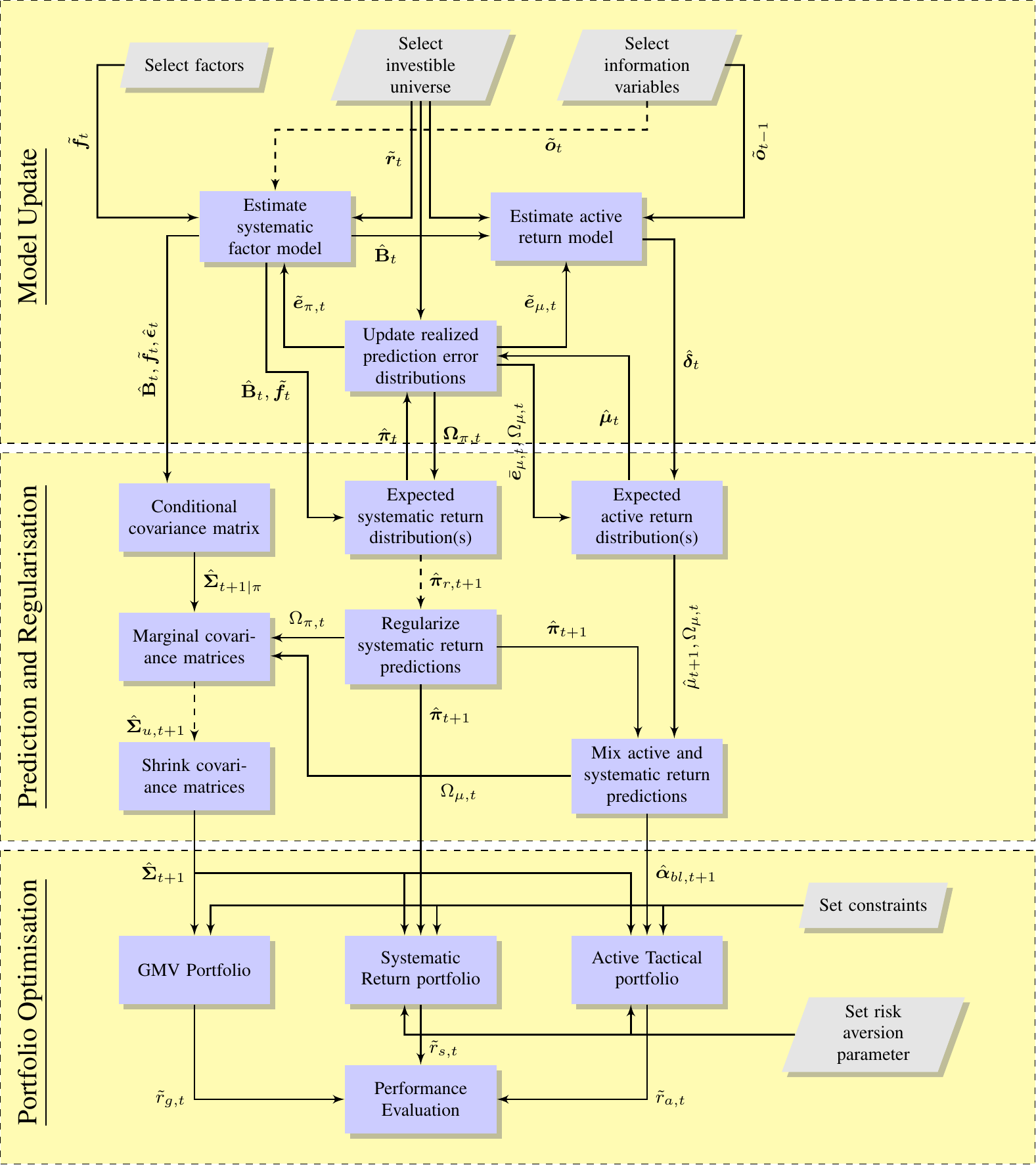}
	\caption[Annotated Flow Chart for a Quantitative Investment Workflow]{Annotated Flow Chart for a Quantitative Investment Workflow: The flow chart represents a simplified process for making repeatable online investment decisions. Blue blocks represent online tasks and grey blocks represent offline tasks or decisions, whilst arrows represent the flow of data and variables between tasks.  At each time step, new data becomes available, models are updated, predictions are prepared, and portfolios are constructed. Model updates are performed online by feeding back prediction errors ($\tilde{e}_{\pi,t}$ and $\tilde{e}_{\mu,t}$) to correct parameter estimates. The dotted arrows represent information flows that were not implemented in Section \ref{sec:results}. The same chart is shown in Figure \ref{fig:workflow_symbols} except with formulae in place of descriptive labels. Nomenclature is provided in Table \ref{tbl:Nomenclature}.}\label{fig:workflow_words}
\end{figure}


\newpage
\onecolumn
\subsection{Theoretical Concepts Workflow} \label{app:theoryworkflow}
\begin{figure}[hbtp!]
	
	\centering
	\includegraphics*[width=0.9\textwidth]{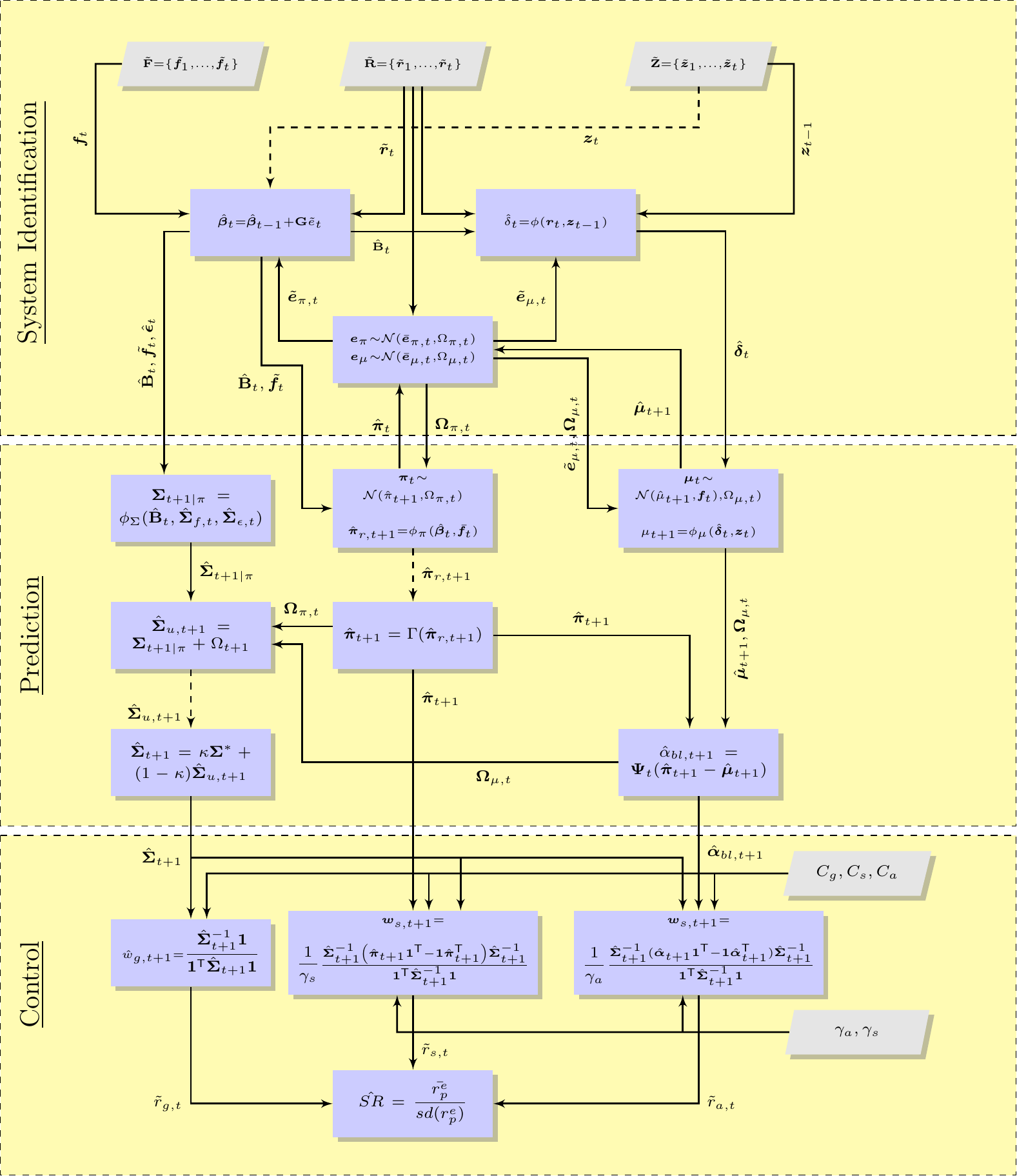}
	\caption[Annotated Flow Chart for a Quantitative Investment Workflow]{Annotated Flow Chart for a Quantitative Investment Workflow: The flow chart represents one iteration of the automated portfolio construction process. The chart reads from top to bottom with arrows representing the flow of data and variables between process components and decisions (blocks). The equations in each block describe how inputs are transformed into outputs to be used at subsequent stages of the investment process. By linking equations, the flow chart depicts how data is transformed online into investment decisions in a serial manner. Linking data to decisions allows for \textit{process-level} analysis of the investment process and is a necessary step for machine learning approaches. The same chart is shown in Figure \ref{fig:workflow_words} except with descriptive labels in place of formulae.  Nomenclature is provided in Table \ref{tbl:Nomenclature} in Appendix \ref{app:nomenclature}}\label{fig:workflow_symbols}
\end{figure}


\end{document}